\documentclass[a4paper,journal]{IEEEtran}
\usepackage{amsmath,amsfonts}
\usepackage{amssymb,amstext}
\usepackage{array}
\usepackage{subfigure}
\usepackage{textcomp}
\usepackage{url}
\usepackage{verbatim}
\usepackage{graphicx}
\usepackage{cite}
\usepackage{multirow,multicol}
\usepackage{psfrag}
\usepackage{rsfs}
\usepackage{jumoline}
\usepackage{ulem}
\usepackage{losymbol}
\usepackage{threeparttable}
\usepackage{flushend}

\renewcommand{\j}{\mathrm{j}}
\newcommand{\R}{\mathrm{R}}
\newcommand{\T}{\mathrm{T}}

\newcommand{\Vect}[1]{\boldsymbol{#1}}

\newcommand{\Ang}[1]{${#1}^{\circ}$}

\begin{document}

\title{Device-Free Localization Using Multi-Link MIMO Channels in Distributed Antenna Networks}

\author{Minseok Kim,~\IEEEmembership{Senior member, IEEE,} Gesi Teng, Keita Nishi, Togo Ikegami, Masamune Sato
%\thanks{Manuscript received April 19, 2021; revised August 16, 2021.}
\thanks{This research has been partially supported by ``Nuclear Energy S\&T and Human Resource Development Project (JPJA22P22683407)'' made with the Japan Atomic Energy Agency (JAEA), and ``R\&D for Expansion of Radio Wave Resources (JPJ000254)'' made with the Ministry of Internal affairs and Communications (MIC), Japan.}
\thanks{Minseok Kim, Gesi Teng and Keita Nish are with the Graduate School of Science and Technology, Niigata University, Niigata 950-2181, Japan (e-mail: mskim@ieee.org).}
\thanks{Togo Ikegami is with Honda Motor Company Ltd., Tokyo 107-8556, Japan.}
\thanks{Masamune Sato is with JUKI industrial Equipment Technology Corp., Akita 019-0701, Japan.}
}

% The paper headers
%\markboth{Journal of \LaTeX\ Class Files,~Vol.~14, No.~8, August~2021}%
%{Shell \MakeLowercase{\textit{et al.}}: A Sample Article Using IEEEtran.cls for IEEE Journals}

%\IEEEpubid{0000--0000/00\$00.00~\copyright~2021 IEEE}
% Remember, if you use this you must call \IEEEpubidadjcol in the second
% column for its text to clear the IEEEpubid mark.

\maketitle

\begin{abstract}
Targeting integrated sensing and communication (ISAC) in future 6G radio access networks (RANs), this paper presents a novel device-free localization (DFL) framework based on distributed antenna networks (DANs). In the proposed approach, radio tomographic imaging (RTI) leverages the spatial and temporal diversity of multi-link multiple-input multiple-output (MIMO) channels in DANs to achieve accurate localization. Furthermore, a prototype system was developed using software-defined radios (SDRs) operating in the sub-6 GHz band, and comprehensive evaluations were conducted under indoor conditions involving varying node densities and target types. The results demonstrate that the framework achieves sub-meter localization accuracy in most scenarios and maintains robust performance under complex multipath environments. In addition, the use of Bayesian optimization to fine-tune key parameters, such as sparsity and path thickness, led to significant improvements in image reconstruction quality and target estimation accuracy. These results demonstrate the feasibility and effectiveness of DAN-based DFL as a scalable and infrastructure-compatible ISAC solution, capable of delivering accurate, passive localization without dedicated sensing hardware.
\end{abstract}

\begin{IEEEkeywords}
ISAC, device-free localization, radio tomographic imaging, distributed antenna networks, multi-link channel, multistatic radar, beamforming, distributed MIMO
\end{IEEEkeywords}

\section{Introduction}
\IEEEPARstart{D}{istributed} antenna networks (DANs), such as distributed massive multiple-input-multiple-output (D-MIMO) \cite{distributed_MIMO} and cell-free MIMO \cite{cell_free}, have recently gained prominence as foundational technologies for next-generation wireless communication systems. By deploying a large number of antennas over a wide area rather than concentrating them in a single array, these systems achieve substantial improvements not only in network capacity and reliability but also in localization accuracy and sensing capabilities. The expansion in the number of antennas increases the spatial degrees of freedom, leading to improved spectral efficiency and diversity, while also reducing the impact of large-scale fading. Furthermore, the inherently distributed nature of these systems facilitates the utilization of multistatic radar information, thereby enabling integrated sensing and communication (ISAC), which is recognized as a key application area in the development of 6G. In 6G DAN-based ISAC systems, localization and position estimation are expected to serve as key enabling technologies \cite{Guo,Demirhan}.

ISAC utilizes communication waveforms not only for data transmission but also for environmental sensing, enabling high-accuracy localization through techniques such as time of arrival (ToA), time difference of arrival (TDoA), angle of arrival (AoA), and received signal strength (RSS) \cite{Witrisal,Wang}. These capabilities are further enhanced by MIMO orthogonal frequency division multiplexing (OFDM) systems, which offer improved spatiotemporal resolution. In this context, a DAN-based ISAC system can implement multistatic sensing to effectively detect nearby obstacles and human presence. This dual-purpose operation enhances the robustness and reliability of wireless communication links while simultaneously generating valuable sensing outputs. These outputs can be further exploited for device-free localization (DFL), which is particularly useful in various daily-life applications. Unlike vision-based localization, radio-based DFL techniques preserve user privacy and can penetrate non-transparent obstacles such as walls. However, pointwise radio-based DFL methods which estimate target position based on the angle and range of reflected signals are susceptible to performance degradation in multipath-rich environments \cite{Huang, Haimovich, Schmidhammer}. By combining distributed antenna deployment with multistatic sensing, the DAN-based ISAC architecture is expected to mitigate these challenges and significantly improve localization and communication performance in complex non-line-of-sight (NLoS) scenarios.

As an alternative to pointwise methods, radio tomographic imaging (RTI) has been introduced to estimate spatial variations in path gain caused by target-induced shadowing effects. RTI constructs an image of the attenuation field by leveraging dense wireless links within a radio sensor network, thereby enabling DFL and target tracking \cite{Wilson, Related_Bocca, Related_Denis_1, Related_Denis_2, Related_Yang, Related_Ma, Zhang, Related_Tan, Related_Cimdins, Related_Cimdins_2, EuCAP_Kim, Access_Ikegami}. RTI techniques uniformly divide the area of interest into multiple voxels---either as a 2-D grid or a 3-D volumetric mesh---and estimate voxel values from the measured RSS by solving an ill-posed inverse problem. These voxel values represent relative measures, where higher values indicate a greater likelihood of the target being present in the corresponding voxel. When applied in narrowband radio systems such as ZigBee, Bluetooth low energy (LE), or WiFi, accurate RSS measurements are often compromised by multipath fading, which significantly degrades the image quality. Consequently, a large number of antennas (anchor nodes) are typically required to achieve a desired level of spatial resolution; otherwise, substantial performance degradation is observed in multipath environments.
	
Several studies have investigated RTI techniques that are tolerant of or assisted by multipath propagation \cite{Related_Ma, Zhang, Related_Tan, Related_Cimdins, Related_Cimdins_2, EuCAP_Kim, Access_Ikegami}. In \cite{Related_Ma,Zhang}, a DFL method utilizing phase response shift (PRS) was proposed to mitigate the effects of multipath and noise without introducing phase ambiguity. Further, the feasibility of applying static reflected multipath components to DFL is explored. In \cite{Related_Tan}, low-rank and sparse decomposition techniques are employed to remove image artifacts caused by multipath, thereby improving RTI image accuracy. To further enhance localization accuracy by leveraging multipath propagation, the studies in \cite{Related_Cimdins, Related_Cimdins_2} proposed multipath-assisted RTI, which extracts multipath components (MPCs) from the channel impulse response (CIR) of ultra-wideband signals, estimates single-bounce reflection paths, and incorporates them into the RTI framework. Although this approach enables rapid localization with a small number of anchor nodes, its accuracy is often limited due to the lack of angle information of MPCs. On the other hand, multipath-RTI, previously proposed by the authors \cite{EuCAP_Kim, Access_Ikegami}, was developed for use with millimeter-wave (mm-wave) radio systems such as 5G new radio access network (NR) and WiGig WLAN, which are currently attracting considerable attention due to their high-resolution channel acquisition capabilities in both the delay and angle domains. In this approach, the RSS variation of each MPC can be individually exploited, enabling the use of a large number of additional virtual anchor nodes. As a result, the system performance can be improved by reducing the number of physical anchor nodes through the utilization of multipath propagation. 

Whereas the multipath-RTI offers advantages in multipath-rich environments, it requires accurate information about the physical pathways of MPCs. This implies that effective separation of MPCs remains a major challenge, as existing super-resolution path parameter estimation techniques are computationally intensive and complex \cite{Fleury_SAGE,Richter,subgridclean}. In addition, identifying propagation pathways, at least up to double-bounce reflections, requires precise clustering of MPCs, a process hindered by the limited spatiotemporal resolution of practical systems. Furthermore, determining the physical pathways corresponding to each MPC (or cluster) is technically demanding; therefore, accurate measurement-based ray tracers (MBRTs) \cite{Poutanen} and automated path-association methods that match propagation paths between current and baseline measurements must be needed. To address the aforementioned problem, the study in \cite{Access_Ikegami} investigates a ray-tracing-assisted multipath-RTI approach. Ray-tracing (RT) will enable real-time channel prediction on graphics processing units (GPUs) by leveraging environmental geometry obtained through LiDAR, and is expected to play a pivotal component in future 6G ISAC systems. In this method, the propagation paths of individual MPCs are predicted based on a geometric model of the surrounding environment, eliminating the need for exhaustive signal processing of measurement data. Experimental evaluations demonstrated that this approach achieves high localization accuracy, with positioning errors below $0.5$ meters. However, due to the ray-optical nature of mm-wave signal propagation, the coverage is inherently limited, even though the significant attenuation caused by blockages can contribute to enhanced RTI performance within confined areas. As a result, a large number of anchor nodes must be deployed to ensure sufficient spatial coverage.

To address the challenges outlined above, this paper extends the prior work \cite{Access_Ikegami} by developing a DFL technique based on DANs. Specifically, a prototype system has been realized, featuring a sub-6 GHz multi-link channel sounding architecture for multistatic radar applications. The system is implemented using software-defined radios (SDRs)~\cite{Ettus}, enabling flexible and reconfigurable signal processing. The proposed system employs a switched-antenna scheme to support a large number of antennas at a single distributed unit and enables accurate multistatic radar measurements by synchronizing multiple SDR nodes. These advancements offer methodological and experimental improvements, and establish a practical, high-performance foundation for advancing DFL in DAN-based ISAC systems. Importantly, DANs envisioned as key infrastructure in future 6G ISAC systems are inherently designed for synchronized operation, both within individual antenna arrays and across multiple spatially distributed nodes, leveraging shared timing references such as fiber-optic distribution, GNSS-disciplined oscillators, or time-sensitive networking protocols. Furthermore, environmental geometry is increasingly available through auxiliary technologies as mentioned above. The proposed multipath-RTI framework is therefore built upon capabilities that are expected to be native to next-generation DAN-based deployments. While this approach introduces additional complexity compared to conventional RSS-based RTI, it remains practically feasible within synchronized, infrastructure-supported networks, as targeted in emerging 6G ISAC scenarios.

The main contributions of this study are as follows:
\begin{enumerate}
\item
%DAN-DFLアルゴリズム
A DFL technique leveraging multi-link MIMO channels in DANs is developed. Assuming that each distributed antenna unit is equipped with an antenna array and that wireless data transmission adopts a MIMO-OFDM scheme, this paper formulates multipath-RTI localization using double-directional spatiotemporal beamforming.

\item
%DAN-DFLシステム実装
Based on the feasibility study presented in \cite{Access_Ikegami}, this paper develops a four-node DAN-based DFL prototype system operating in the sub-6 GHz band to demonstrate its practical validity. The system incorporates a highly flexible and versatile multi-link MIMO channel sounding architecture for multistatic radar-based target localization, implemented using SDRs.

\item
%DAN-DFL実験による評価
Using the developed system, this paper presents a measurement-based performance evaluation of position estimation accuracy in an indoor test environment. The evaluation provides insight into the practical feasibility of a DAN-based DFL, highlighting the system's ability to achieve sub-meter accuracy under multipath-rich conditions.

\end{enumerate}

The remainder of this paper is organized as follows: Section~II introduces the proposed DFL framework based on multipath-RTI in DANs. Section~III presents simulation-based evaluations, analyzing the effects of node count and antenna array size on localization accuracy. Section~IV describes the prototype system implementation using SDRs, highlighting a switched-antenna MIMO channel sounding scheme and a multi-link channel acquisition architecture for DANs. Section~V provides measurement-based results under indoor conditions, including analysis of node count and the impact of different target types. Section~VI concludes the paper.

\section{DAN-based DFL}
\subsection{System Model}
\begin{figure}[t]
\centerline{
\includegraphics[width=0.99\linewidth]{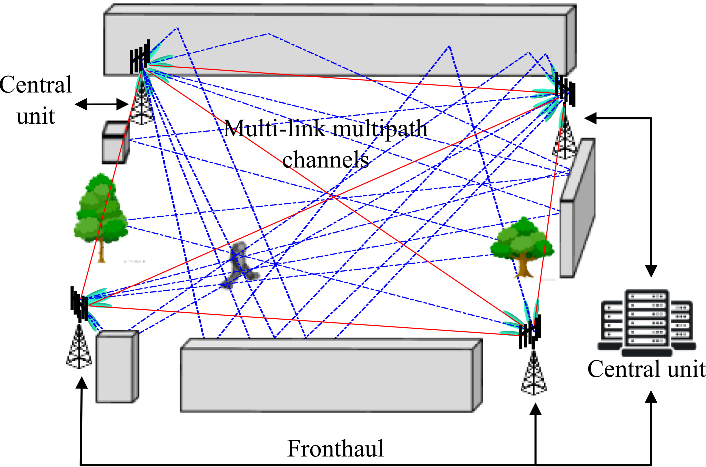}}
\caption{DFL utilizing multi-link multipath channels in a DAN. \label{fig:DFL_DAN}}
\end{figure}

Fig.~\ref{fig:DFL_DAN} illustrates the system model for DFL using multi-link multipath channels in a DAN. The DAN architecture comprises a central unit (CU) and a group of distributed antenna units, such as remote radio heads (RRHs), deployed throughout the coverage area. These distributed antenna units are connected to the CU via dedicated fronthaul links. In this configuration, the instantaneous RSS of each MPC, extracted at the distributed antenna units through beamforming, is transmitted to the CU over the fronthaul network. The CU then performs centralized signal processing on the aggregated RSSs to enable ISAC DFL. It is assumed that each distributed antenna unit is equipped with an antenna array, and that wireless data transmission is based on a MIMO-OFDM scheme with a bandwidth of $W$. The wireless links among distributed antenna units, functioning as anchor nodes, consist of numerous MPCs arising from reflection, diffraction, and penetration, in addition to the LoS path. Notably, reflected paths can be interpreted as originating from virtual anchor nodes located outside the physical deployment area. By leveraging these virtual anchor nodes, it is possible to obtain a sufficient number of effective anchors for DFL, even when only a limited number of physical anchor nodes are available.

\subsection{DFL using Multipath-RTI}
Consider the RSS measurements across $N$ links (pathways) in a two-dimensional area of interest, which is uniformly divided into $M$ voxels. The measured RSS values (usually in decibel), $\Vect{y}(t) \in \mathbb{R}^N$ for each MPC are represented as a linear combination of voxel values along their respective propagation paths as
\begin{equation}
\Vect{y}(t)=\Vect{W} \Vect{x}(t)+\Vect{n}(t),
\end{equation}
where $\Vect{x}(t)\in\mathbb{R}^M$ represents the voxel values in the target localization area, and $\Vect{n}(t) \in \mathbb{R}^N$ is measurement noise. The change in RSS due to power attenuation caused by obstruction at time $t$ is,
\begin{equation}
\Delta \Vect{y}(t)=\Vect{W} \Delta \Vect{x}(t)+\Delta \Vect{n}(t),
\label{eq:deltay}
\end{equation}
where $\Delta \Vect{y}(t)=\Vect{y}(t)-\Vect{y}_0$, $\Delta \Vect{x}(t)=\Vect{x}(t)-\Vect{x}_0$, and $\Delta \Vect{n}(t)=\Vect{n}(t)-\Vect{n}_0$. In addition, $\Vect{y}_0 = \Vect{W} \Vect{x}_0 + \Vect{n}_0$ denotes the values in the baseline measurement. Shadowing effects caused by obstructions result in multiplicative changes in signal power, which become subtractive when expressed in the logarithmic (dB) domain as \eqref{eq:deltay}. Consequently, changes in RSS, $\Delta \Vect{y}(t)$, can be modeled as a linear combination of voxel-level attenuations.
The transformation matrix $\Vect{W} \in \mathbb{R}^{N \times M}$ maps the sparse vector $\Delta \Vect{x}(t)$ to the corresponding observation $\Delta \Vect{y}(t)$. Each entry of $\Vect{W}$ is determined by assigning appropriate weights to the voxels that lie along predetermined propagation paths, as proposed in \cite{Wilson}. Namely, the matrix $\boldsymbol{W}$ encapsulates the multipath information under baseline (static) conditions. It is assumed that the structure of $\boldsymbol{W}$ remains unchanged over time, as long as no new propagation paths emerge and no existing ones vanish. It is expected that only the power levels along these paths vary in response to dynamic obstructions such as human movement.

%%%%%%%%%%%%%%%%%%%%%%%%%%%%%%%%%%%%%%%%%%%%%%%%%%%%%%%%%%%%%%%%%%%%
%%%%%%%%%%%%%%%%%%%%%%%%%%%%%%%%%%%%%%%%%%%%%%%%%%%%%%%%%%%%%%%%%%%%

\begin{figure}[t]
\centerline{
\includegraphics[width=0.99\linewidth]{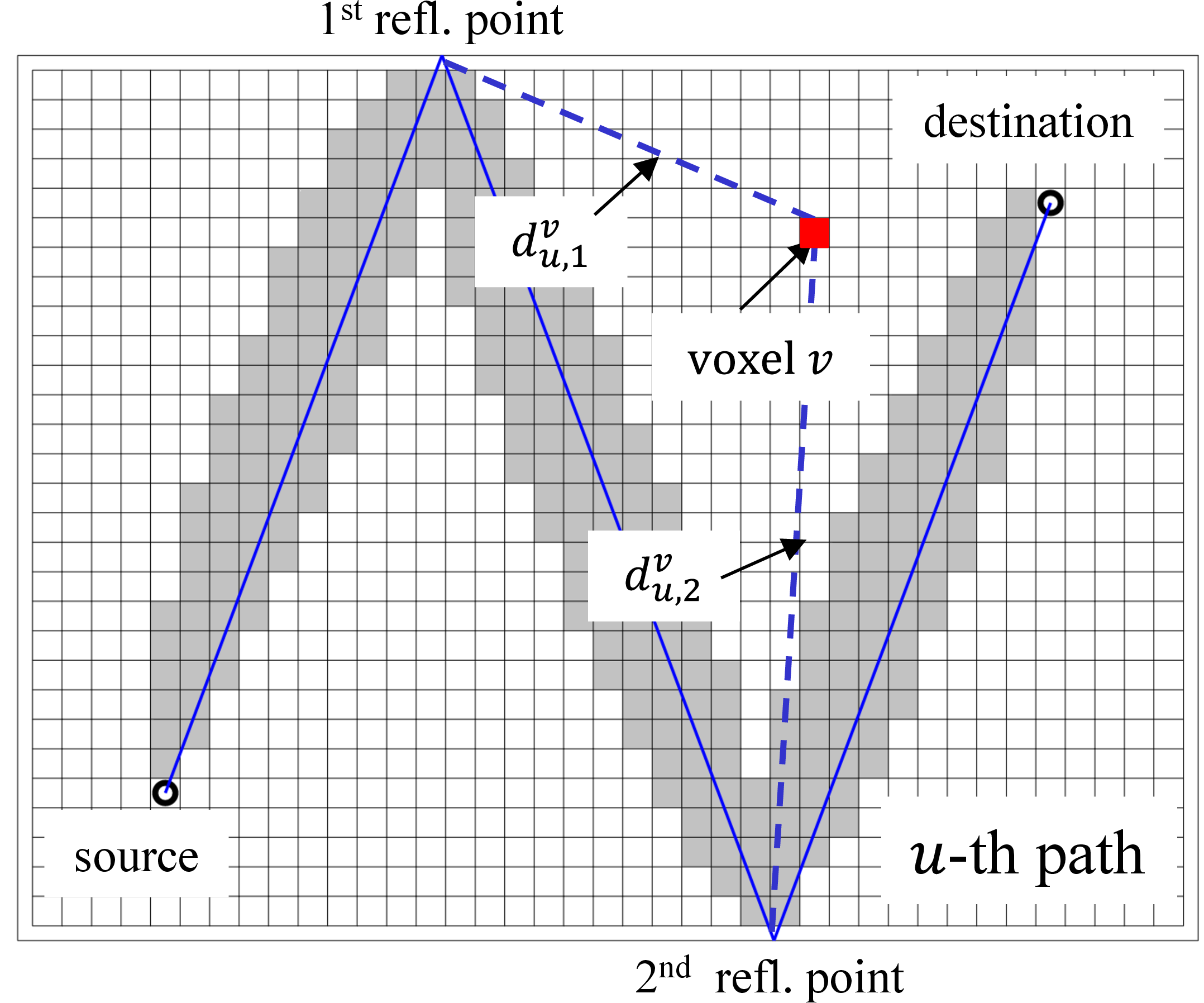}}
\caption{Determination of non-zero voxel elements in the weight matrix corresponding to the propagation pathway between two arbitrary nodes in a DAN. \label{fig:Weighting}}
\end{figure}

It is assumed that the DAN consists of $N_\mathrm{node}$ nodes, with the total number of links, given by $L={N_\mathrm{node} \choose 2}$, and that the transmitter (Tx) and receiver (Rx) at each end of the $l$-th link are equipped with $M_{\mathrm{T},l}$ and $M_{\mathrm{R},l}$ antenna elements, respectively. Further, the total number of paths $N=\sum_{l=1}^{L} N_{l}$ where $N_{l}$ denotes the number of propagation paths associated with the $l$-th link. As shown in Fig.~\ref{fig:Weighting}, $\Vect{W}$ is determined as \cite{Access_Ikegami}
\begin{eqnarray}
[\Vect{W}]_ {u,v} = \left\{\begin{array}{ll}{1/\sqrt{d_{u}},}&{d_{u,\mathrm{src}}^{v}+d_{u,1}^{v}<d_{u,1}+\gamma} \\{1/\sqrt{d_{u}},}&{d_{u,1}^{v} \ \ +d_{u,2}^{v}<d_{u,2}+\gamma} \\ &\vdots \\{1/\sqrt{d_{u}},}&{d_{u,{K_u}}^{v}+d_{u,\mathrm{des}}^{v}<d_{u,K_u+1}+\gamma} \\ {0,}&{\text{elsewhere}}\end{array}\right.
\label{eq:weightmatrix3}
\end{eqnarray}
Here, let $u = \sum_{k=1}^{l-1} N_{k} + i$ be the path serial index corresponding to the $i$-th path of the $l$-th link. $d_u = \sum_{k=1}^{K_u+1} d_{u,k} $ represents the total travel distance along the $u$-th pathway. The terms $d_{u,\mathrm{src}}^{v}$, $d_{u,r_{u}}^{v}$, and $d_{u,\mathrm{des}}^{v}$ denote the distances between the $v$-th voxel and the source node, the $r_{u}$-th reflecting point, and the destination node, respectively, where $r_{u} \in \left\{ \mathrm{src}, 1, \cdots, K_u, \mathrm{des} \right\}$. $K_u$ denotes the number of reflections of the $u$-th path. Furthermore, $\gamma$ is a margin parameter that affects the thickness of a pathway. The propagation paths of higher-order reflections tend to become increasingly complex, making them impractical to utilize in real-world environments. As the reflection order increases, the pathways become less correlated with the actual target position, leading to a higher likelihood of generating artifacts. Therefore, utilizing up to second-order reflections is a reasonable choice. 

Since $\Vect{W}$ becomes a sparse matrix, the change in voxels, $\Delta \Vect{x}$ in~\eqref{eq:deltay}, can be obtained by solving an ill-posed inverse problem using various optimization techniques such as least squares, compressed sensing, and so forth. 
Regularization by Elastic Net \cite{ElasticNet}, a hybrid approach that combines $L_1$ and $L_2$ penalties, is employed to estimate $\Delta\hat{\Vect{x}}$, the reconstruction of $\Delta \Vect{x}$ from the observation $\Delta \Vect{y}$ as
\begin{eqnarray}
\Delta\hat{\Vect{x}} = \arg\min_{\Delta\Vect{x}} \left\{ \frac{1}{2} \left\| \Delta\Vect{y} - \Vect{W} \Delta\Vect{x} \right\|_2^2 + \lambda P_{\alpha}(\Delta\Vect{x}) \right\},
\label{eq:ElasticNet}
\end{eqnarray}
where the Elastic Net penalty term is defined as
\begin{eqnarray}
P_\alpha(\Delta\Vect{x})=\frac{1-\alpha}{2}\|\Delta\Vect{x}\|_2^2+\alpha\|\Delta\Vect{x}\|_1.
\end{eqnarray}
$\lambda$ and $\alpha$ control regularization intensity and image sparsity, respectively \cite{Access_Ikegami}. 
Here, $\lambda$ controls the overall regularization strength, while $\alpha \in [0, 1]$ balances the contribution between the $L_1$ (sparsity-promoting) and $L_2$ (stability-enhancing) terms~\cite{Access_Ikegami}. 
The actual position within the area of interest where the RSS change occurs can be specified based on the relationship between the physical target location and the corresponding voxel. To locate the target position, a binary image is first generated by binarizing $\Delta \Vect{x}$, assigning a value of 1 to voxels deemed active. A clustering algorithm, such as density-based spatial clustering of applications with noise (DBSCAN) \cite{DBSCAN}, is then applied to group spatially adjacent active voxels. The centroids of the dominant clusters are subsequently selected as the final position estimate.
As described above, striking an appropriate balance between the $L_1$ and $L_2$ terms is especially important for achieving high-quality image reconstruction that enables accurate and reliable localization. Elastic Net not only accommodates spatial discontinuities by promoting both sparsity and stability, but also integrates well with the post-processing step based on DBSCAN clustering. Specifically, it serves as an initial sparse selector that identifies candidate regions for localization, which are subsequently refined through clustering. While precise tuning of the regularization parameters, which control the trade-off between sparsity and stability, is essential for optimal performance, this tuning process is highly challenging in practice. Alternative strategies may offer further improvements in image reconstruction quality and localization accuracy.

%%%%%%%%%%%%%%%%%%%%%%%%%%%%%%%%%%%%%%%%%%%%%%%%%%%%%%%%%%%%%%%%%%%%
%%%%%%%%%%%%%%%%%%%%%%%%%%%%%%%%%%%%%%%%%%%%%%%%%%%%%%%%%%%%%%%%%%%%

\subsection{Signal Processing in DANs}
The CIR, sampled at delay $\check{\tau}_{d}$ using $M_{\T,l}$ and $M_{\R,l}$ antenna elements at each end of the $l$-th link, is expressed in the form of a MIMO channel matrix through the superposition of plane wave MPCs as
\begin{equation}
\Vect{H}_l(\check{\tau}_{d}) = \sum_{i=1}^{N_{l}} \gamma_{l,i} \, a_\tau(\check{\tau}_{d}-\tau_{l,i}) \, \Vect{a}_\R(\phi_{\R,l,i}) \, \Vect{a}_\T^T(\phi_{\T,l,i}),
\label{eq:H_MIMO}
\end{equation}
where $N_{l}$ denotes the number of MPCs for the $l$-th link, $\gamma_{l,i}$ is the complex gain of the $i$-th path, $a_\tau(\cdot)$ is the signal autocorrelation function, and $\Vect{a}_\T(\cdot)$ and $\Vect{a}_\R(\cdot)$ are the array response vectors at the Tx and Rx respectively. $\check{\tau}_{d}$ represents the $d$-th delay bin, given by $d \cdot W^{-1}$ ($d=0,...,D-1$), where $W$ denotes the signal bandwidth. The parameters $\tau_{l,i}$, $\phi_{\T,l,i}$, and $\phi_{\R,l,i}$ denote the delay, azimuth angle of departure, and azimuth angle of arrival of the $i$-th MPC associated with the $l$-th link, respectively.

The CIR in the vectorized form of \eqref{eq:H_MIMO} for the $l$-th link across $D$ delay samples is expressed as
\begin{equation}
\Vect{h}_l = \Vect{A}(\Vect{\Omega}_{l})\, \Vect{\Gamma}_l \in \mathbb{C}^{M_{\T,l} M_{\R,l} D \times 1},
\label{eq:h_vectorized}
\end{equation}
where $\Vect{A}(\Vect{\Omega}_l)$ is the response matrix constructed from the delay-angular domain parameters $\Vect{\Omega}_l$, defined as
\begin{equation}
\Vect{A}(\Vect{\Omega}_l) = \left[\Vect{a}\left(\Vect{\Omega}_{l,1}\right), \ldots, \Vect{a}\left(\Vect{\Omega}_{l,N_{l}}\right)\right] \in \mathbb{C}^{M_{\T,l} M_{\R,l} D \times N_{l}},
\label{eq:A_definition}
\end{equation}
where each $\Vect{a}\left(\Vect{\Omega}_{l,n_{l}}\right)$ corresponds to the combined response associated with the $n_{l}$-th MPC. The vector $\Vect{\Gamma}_l$ contains the complex path weights of the MPCs and is expressed as
\begin{equation}
\Vect{\Gamma}_l = \left[\gamma_{l,1}, \ldots, \gamma_{l,N_{l}}\right]^T \in \mathbb{C}^{N_{l} \times 1}.
\label{eq:Gamma_definition}
\end{equation}
Each delay-angular domain parameter vector is defined as
\begin{equation}
\Vect{\Omega}_{l,n_{l}} = \left[\tau_{l,n_{l}}, \phi_{\T,l,n_{l}}, \phi_{\R,l,n_{l}} \right].
\label{eq:Omega_definition}
\end{equation}
Finally, the combined response vector $\Vect{a}\left(\Vect{\Omega}_{l,n_{l}}\right)$ is expressed as the Kronecker product of the Tx and Rx antenna responses, and signal autocorrelation vectors:
\begin{multline}
\Vect{a}\left(\Vect{\Omega}_{l,n_{l}}\right) = \Vect{a}_\T\left(\phi_{\T,l,n_{l}}\right) \otimes \Vect{a}_\R\left(\phi_{\R,l,n_{l}}\right) \otimes \Vect{a}_\tau\left(\tau_{l,n_{l}}\right) \\ \in \mathbb{C}^{M_{\T,l} M_{\R,l} D \times 1}.
\label{eq:a_definition}
\end{multline}
For simplicity, we consider uniform linear arrays (ULAs) with half-wavelength spacing, where each antenna element exhibits an omnidirectional radiation pattern. The corresponding steering vectors are given by 
\begin{equation}
\Vect{a}_{\mathrm{T}}\left(\phi_{\T}\right)
=\left[1, e^{-\j \pi \sin \phi_{\T}}, \ldots, e^{-\j \pi(M_{\T,l}-1) \sin \phi_{\T}}\right]^T,
\label{eq:a_T}
\end{equation}
\begin{equation}
\Vect{a}_{\mathrm{R}}\left(\phi_{\R}\right)
=\left[1, e^{-\j \pi \sin \phi_{\R}}, \ldots, e^{-\j \pi(M_{\R,l}-1) \sin \phi_{\R}}\right]^T,
\label{eq:a_R}
\end{equation}
which can be replaced by the measured antenna manifold. The signal autocorrelation function vector $\Vect{a}_\tau(\tau)$ is defined as
\begin{equation}
\Vect{a}_\tau(\tau) = \left[a_\tau\left(\check{\tau}_0 - \tau\right), \ldots, a_\tau\left(\check{\tau}_{D-1} - \tau\right)\right]^T,
\label{eq:}
\end{equation}
where $a_\tau(\tau)$ denotes the autocorrelation function of an OFDM preamble signal with a rectangular power spectrum (without pulse shaping) and is given \cite{subgridclean} by
\begin{equation}
a_\tau(\tau) = \frac{1}{D} \exp\left(-\mathrm{j} \pi \Delta_f \tau\right) \cdot \frac{\sin\left(\pi D \Delta_f \tau\right)}{\sin\left(\pi \Delta_f \tau\right)},
\label{eq:a_tau_function}
\end{equation}
with $\Delta_f$ representing the subcarrier spacing.

To realize multipath-RTI, it is necessary to estimate the RSS changes associated with each individual propagation path. In this study, as described above, it is assumed that array antennas are deployed at each anchor node to enable separation and extraction of RSS corresponding to distinct multipath components. Leveraging prior knowledge of the propagation paths, $\Vect{\Omega}_l$, the observation vector that aggregates the RSS changes across all multipaths is obtained as
\begin{equation}
\Delta \Vect{y} = \left[\Delta \Vect{y}_1^T, \ldots, \Delta \Vect{y}_L^T\right]^T \in \mathbb{C}^{N \times 1},
\label{eq:delta_y}
\end{equation}
where $N$ represents the total number of multipath components across all $L$ links. The RSS change vector for the $l$-th link, 
\begin{eqnarray}
\Delta \Vect{y}_l = \Vect{y}_l(t) - \Vect{y}_{\mathrm{base},l}(t) \ \mathrm{[dB]},
\label{eq:delta_yl}
\end{eqnarray}
where $\Vect{y}_l$ and $\Vect{y}_{\mathrm{base},l}$, are computed via spatiotemporal beamforming. The $n_l$-th element of \eqref{eq:delta_yl} is calculated as
\begin{multline}
[\Delta \Vect{y}_l]_{n_l} = 10\log_{10}\left(\frac{\mathbb{E}\left[|\Vect{a}^H(\Vect{\Omega}_{l,n_l}) \Vect{h}_{l}(t)|^2\right]}{\mathbb{E}\left[|\Vect{a}^H(\Vect{\Omega}_{l,n_l}) \Vect{h}_{\mathrm{base},l}(t)|^2\right]}\right) \\
 = 10\log_{10} \left( \frac{\Vect{a}^H(\Vect{\Omega}_{l,n_l}) \Vect{R}_{l}\Vect{R}_{l}^H\Vect{a}(\Vect{\Omega}_{l,n_l})}{\Vect{a}^H(\Vect{\Omega}_{l,n_l})\Vect{R}_{\mathrm{base},l}\Vect{R}^H_{\mathrm{base},l} \Vect{a}(\Vect{\Omega}_{l,n_l})} \right),
\label{eq:delta_yln}
\end{multline}
where $\Vect{a}(\Vect{\Omega}_{l,n_l})$ denotes the delay-angular response function vector as defined in~\eqref{eq:a_definition}, and $\Vect{h}_l(t)$ and $\Vect{h}_{\mathrm{base},l}(t)$ denote the MIMO channel vectors measured at time $t$ and measured in the absence of dynamic targets, respectively, for the $l$-th link. The matrices $\Vect{R}_l$ and $\Vect{R}_{\mathrm{base},l}$ are correlation matrices computed as
\begin{align}
\Vect{R}_{l} &= \mathbb{E}\left[ \Vect{h}_l(t) \Vect{h}_l^H(t)\right],
\label{eq:R_l} \\
\Vect{R}_{\mathrm{base},l} &= \mathbb{E}\left[\Vect{h}_{\mathrm{base},l}(t) \Vect{h}_{\mathrm{base},l}^H(t)\right],
\label{eq:R_base_l}
\end{align}
where $\mathbb{E}[\cdot]$ denotes the expectation operator, which is estimated by averaging multiple snapshots over a short duration. The set of individual propagation paths for each link, $\Vect{\Omega}_l$, can be pre-estimated via measurements, ray tracing based on a simplified geometric model of the environment, or a combination of both. Indeed, obtaining accurate propagation paths from measurements alone is challenging. Furthermore, although ray tracing is relatively straightforward to implement, its accuracy is often constrained by discrepancies between simulated predictions and real-world observations. To address this, a path-association method can be applied to identify the actual propagation pathways present among the ray-tracing results, thereby enhancing localization accuracy \cite{Access_Ikegami}.

%%%%%%%%%%%%%%%%%%%%%%%%%%%%%%%%%%%%%%%%%%%%%%%%%%%%%%%%%%%%%%%%%%%%
\begin{figure*}[t]
\centering
\subfigure[Positions of nodes and targets.\label{fig:simulation_model}]{\includegraphics[width=0.27\linewidth]{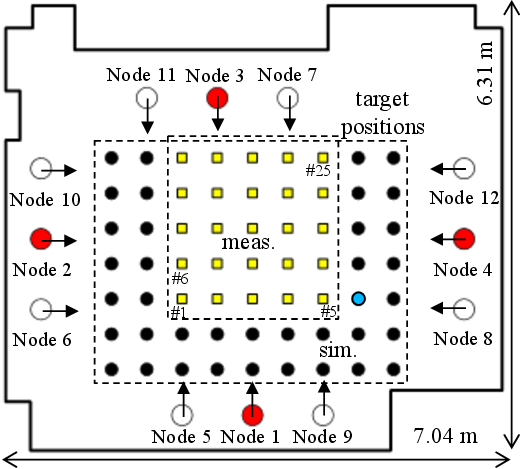}} \qquad    
\subfigure[RT results for a 4-node configuration, displaying propagation paths up to the 2nd-order reflections.\label{fig:simulation_model_rays}]{\includegraphics[width=0.27\linewidth]{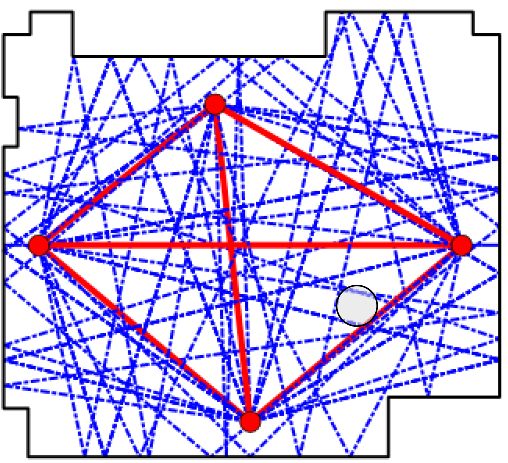}} \qquad
\subfigure[Localization example for a 4-node configuration (error: $0.05$ m).\label{fig:simulation_result_example}]{\includegraphics[width=0.27\linewidth]{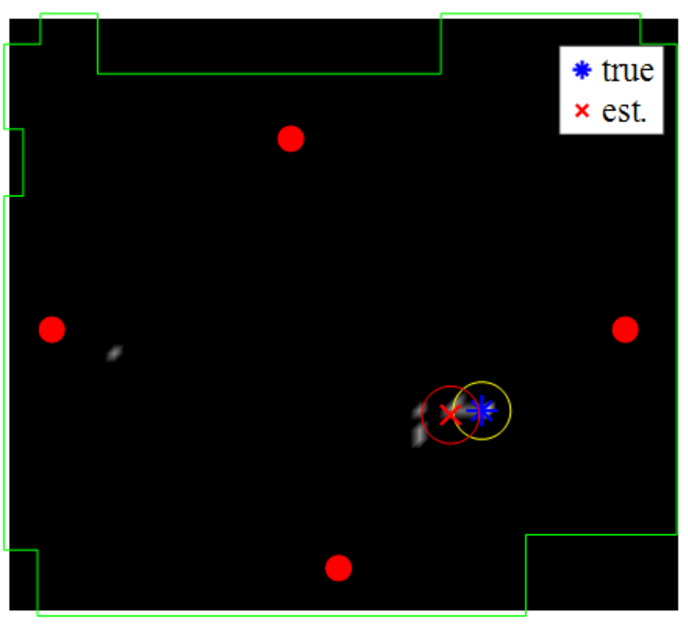}}    
\caption{Evaluation model of an indoor DAN-based DFL system.}
\label{fig:evaluation_setup}
\end{figure*}
%%%%%%%%%%%%%%%%%%%%%%%%%%%%%%%%%%%%%%%%%%%%%%%%%%%%%%%%%%%%%%%%%%%%
%%%%%%%%%%%%%%%%%%%%%%%%%%%%%%%%%%%%%%%%%%%%%%%%%%%%%%%%%%%%%%%%%%%%
\begin{figure*}[t]
\centering
\subfigure[Number of nodes (single antenna per node).\label{fig:cdf_Nodes_1elem}]{\includegraphics[width=0.32\linewidth]{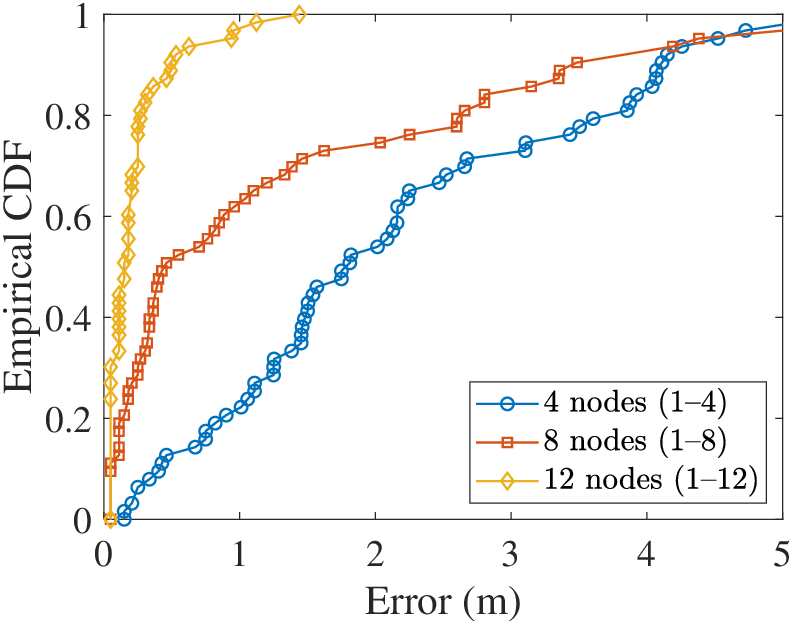}}
\subfigure[Number of nodes (eight elements per node).\label{fig:cdf_Nodes_8elem}]{\includegraphics[width=0.32\linewidth]{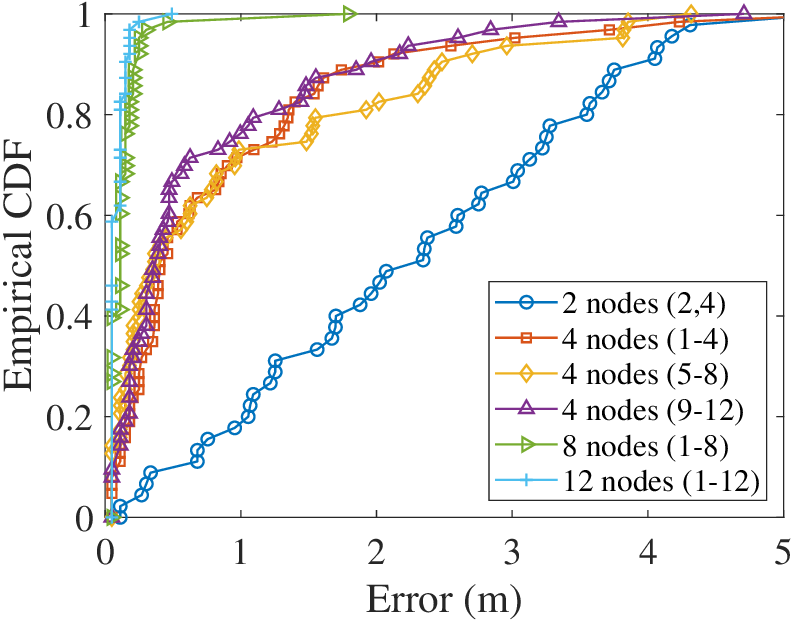}}
\subfigure[Size of antenna array (four nodes).\label{fig:cdf_Elems}]{\includegraphics[width=0.32\linewidth]{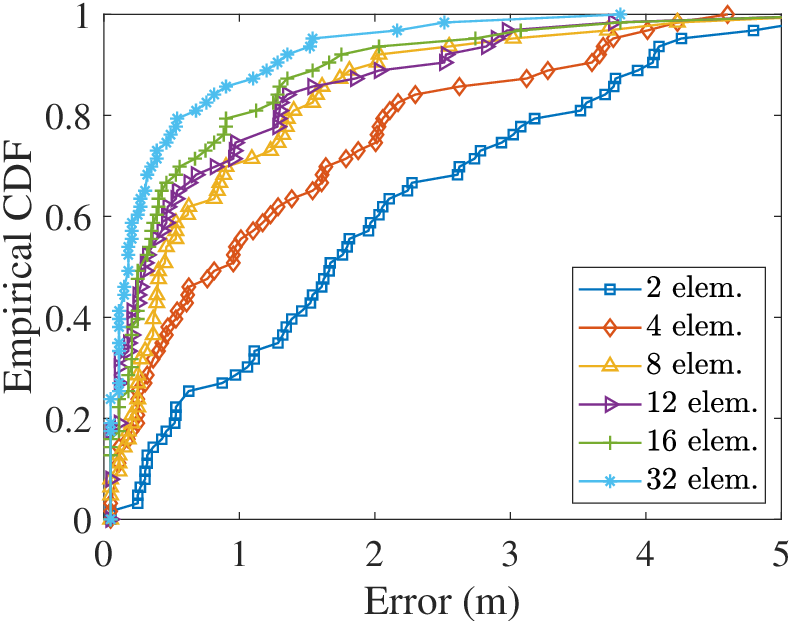}} \quad
\caption{Simulation results.}
\label{fig:SimulationResults}
\end{figure*}
%%%%%%%%%%%%%%%%%%%%%%%%%%%%%%%%%%%%%%%%%%%%%%%%%%%%%%%%%%%%%%%%%%%%

\section{Simulation-Based Optimization of DAN Configurations}

While the preliminaries and parameter design of multipath-RTI have been investigated in \cite{Access_Ikegami}, where simulations idealized the attenuation of individual paths by assuming perfect separation, the impact of antenna array characteristics on inter-path interference must be taken into account to more accurately assess performance in practical propagation environments and to guide proper parameter selection. This section presents simulation results based on the reconstructed multi-link MIMO channels, obtained from RT results using \eqref{eq:H_MIMO}. 

\subsection{Evaluation Setup}

Fig.~\ref{fig:evaluation_setup} illustrates the evaluation model of a four-node indoor DFL system based on a DAN. In Fig.~\ref{fig:simulation_model}, red/white and small black circular markers indicate the anchor node and target locations, respectively. RT simulations were performed for six links in a 3D room model with plasterboard walls and a floor area of $7.04 \times 6.31$ m$^2$. The Tx and Rx antennas for each link were positioned at the same height of $1.3$~m. Fig.~\ref{fig:simulation_model_rays} shows the calcuated ray paths up to double-bounce reflections. A cylinder with a diameter of $0.6$~m and a height of $2$~m was used to model the human target. In the simulation, signal paths obstructed by the target were assumed to be fully shadowed, with no consideration of secondary propagation mechanisms such as diffraction or transmission, in order to simplify the analysis. Assuming that the propagation paths are perfectly known, the vector of RSS changes in \eqref{eq:delta_y} is obtained via double-directional spatiotemporal beamforming, as expressed in \eqref{eq:delta_yln}. The image is subsequently generated from the voxel vector computed using \eqref{eq:ElasticNet} (voxel size: $0.01$ m). After binarization and clustering of the resulting image, the target positions are estimated. Fig.~\ref{fig:simulation_result_example} presents an example of the simulation result. The regularization parameter $\lambda$ was set to $\lambda_{\mathrm{1SE}}$---the largest value for which the 5-fold cross-validation error remains within one standard error (SE) of the minimum---while $\alpha$ was heuristically set to $0.87$. The DBSCAN clustering parameters $\varepsilon$ and $N_\mathrm{minPts}$ were set to $0.5$ and $3$, respectively. The thickness of the propagation paths in \eqref{eq:weightmatrix3} was chosen as $\gamma = 0.03$ \cite{Access_Ikegami}. The results indicate a localization error of $0.05$~m, demonstrating reasonable accuracy even in scenarios with a limited number of paths traversing the target region as shown in Fig.~\ref{fig:simulation_model_rays}.

\subsection{Results}
Based on the described setup, the impact of the number of nodes and the number of antenna elements per node on localization accuracy is evaluated using 63 distinct target positions, uniformly distributed at $0.5$~m intervals, as shown in Fig.~\ref{fig:simulation_model}. Fig.~\ref{fig:SimulationResults} presents the simulation results in terms of the cumulative distribution functions (CDFs) of the localization errors.

\subsubsection{Number of nodes (single antenna)}
Fig.~\ref{fig:cdf_Nodes_1elem} shows the localization accuracy as a function of the number of nodes, where each node is equipped with a single antenna. Since the images are generated solely based on LoS paths among the nodes, an insufficient number of paths, particularly in configurations with 4 and 8 nodes, results in significant degradation in accuracy. The configuration with 12 nodes achieves acceptable localization performance under these conditions.

\subsubsection{Number of nodes (antenna array)}
Fig.~\ref{fig:cdf_Nodes_8elem} illustrates the localization accuracy for varying numbers of nodes, where each node is equipped with eight antenna elements to enable multipath-RTI. As image reconstruction is based on MPCs, including up to double-bounce reflections, a sufficient number of multipaths, particularly in configurations with more than four nodes, yields acceptable accuracy. Additionally, the configuration with two nodes (i.e., a point-to-point setup) exhibits a large localization error, comparable to that observed in the four-node configuration in Fig.\ref{fig:cdf_Nodes_1elem}, where each node has a single antenna.

\subsubsection{Size of antenna array}
Fig.~\ref{fig:cdf_Elems} presents the localization accuracy as a function of the size of the antenna array, with the number of nodes fixed at four. As is well known, increasing the number of antenna elements enhances angular resolution, which improves multipath separation and reduces inter-path interference, thereby improving image quality. Configurations with fewer than four antenna elements result in large errors, whereas those with more than eight elements achieve acceptable accuracy.

It should be noted that the absolute localization accuracy presented above may vary depending on environmental conditions, sensor placement, antenna characteristics, parameter selection, and other system-level factors. Nevertheless, the results presented provide meaningful insights for the practical design of DAN-based DFL systems.

\section{System Implementation}

This section presents the prototype implementation of a multi-link MIMO channel sounding network, developed using SDRs, for the evaluation of DAN-based DFL systems \cite{Radiolab}. The evaluation system should enable the measurement of multistatic radar across multiple spatially distributed antenna nodes. To date, several studies have focused on the development of multi-link MIMO channel sounders for wireless channel characterization. \cite{Kim_ScalableSounder,Konishi_MLSounder} proposed a scalable MIMO channel sounding technique based on a fully parallel transceiver architecture operating at 11 GHz, supporting both directional and multi-link measurements with three eight-antenna modular transceiver units. In contrast, \cite{Sandra} developed a scalable distributed MIMO system using a switched antenna multiplexing architecture based on SDRs for ISAC, incorporating phase-coherent processing and real-time parameter estimation. By leveraging a switched antenna multiplexing scheme, the proposed system enables cost-effective multi-link MIMO channel sounding without requiring a dedicated radio transceiver for each antenna element. The integration of SDRs provides high flexibility in system design, supporting rapid reconfiguration of signal processing functions such as synchronization, calibration, and beamforming. Moreover, the multi-link channel measurements acquired by the system facilitate the development and validation of advanced DFL algorithms that exploit the multipath diversity inherent in distributed deployments.

%%%%%%%%%%%%%%%%%%%%%%%%%%%%%%%%%%%%%%%%%%%%%%%%%%%%%%%%%%%%%%%%%%%%
\begin{figure}[t]
\centering
\includegraphics[width=\linewidth]{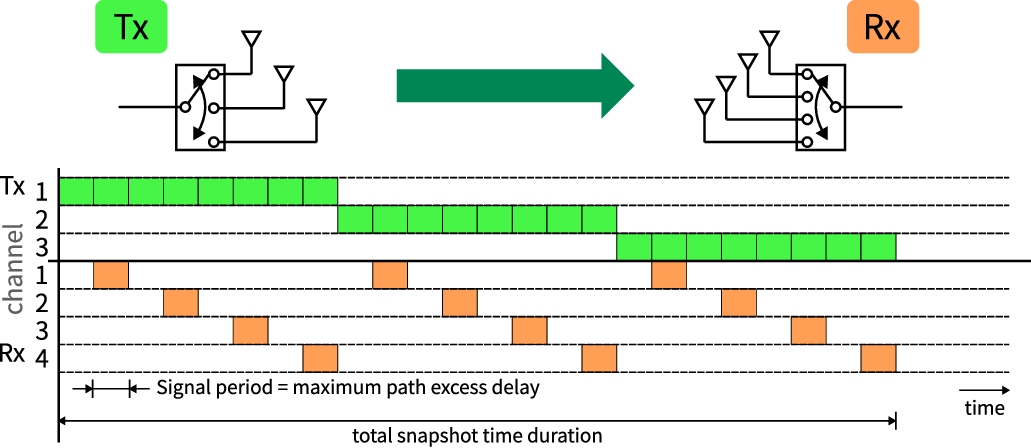}
\caption{Sequence diagram of antenna switching process (example for $3\times 4$ MIMO)}
\label{fig:switching_sequence}
\end{figure}
%%%%%%%%%%%%%%%%%%%%%%%%%%%%%%%%%%%%%%%%%%%%%%%%%%%%%%%%%%%%%%%%%%%%

%%%%%%%%%%%%%%%%%%%%%%%%%%%%%%%%%%%%%%%%%%%%%%%%%%%%%%%%%%%%%%%%%%%%
\begin{table}[t]
\centering
\caption{Specifications of the Switched Antenna $8\times 8$ MIMO Channel Sounder.}
\label{tab:measurement_system}
\begin{tabular}{c|c}
\hline
Item                  & Specification                           \\
\hline
Carrier frequency     & $4.85001$ GHz                           \\
Sampling frequency    & 200 MHz                                 \\
Signal bandwidth      & 100 MHz                                 \\
Sounding signal   & Multitone signal \cite{Kim_ScalableSounder} \\
No. subcarriers & 256                                           \\
Subcarrier interval   & $781.25$ kHz                            \\
Symbol length         & $1.28$ us                               \\
Delay resolution & 10 ns                                        \\
FFT points            & 256                                     \\
Transmit power    & 15 dBm (typ.)                               \\
Measurement time      & 164 us ($8 \times 8$ MIMO)              \\
\hline
\end{tabular}
\end{table}
%%%%%%%%%%%%%%%%%%%%%%%%%%%%%%%%%%%%%%%%%%%%%%%%%%%%%%%%%%%%%%%%%%%%
%%%%%%%%%%%%%%%%%%%%%%%%%%%%%%%%%%%%%%%%%%%%%%%%%%%%%%%%%%%%%%%%%%%%
\begin{figure}[t]
\centering
\includegraphics[width=.8\linewidth]{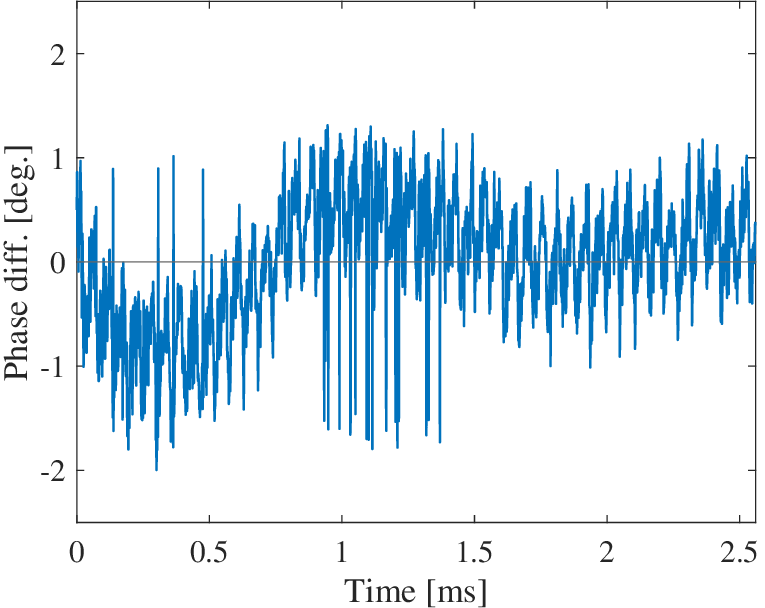}
\caption{Phase stability evaluation.}
\label{fig:phase_stability}
\end{figure}
%%%%%%%%%%%%%%%%%%%%%%%%%%%%%%%%%%%%%%%%%%%%%%%%%%%%%%%%%%%%%%%%%%%%

\subsection{Switched Antenna MIMO Channel Sounding}
First, this subsection elaborates on the point-to-point MIMO channel sounding scheme.

\subsubsection{Switched Antenna Multiplexing Scheme}
To minimize implementation costs, a switched antenna multiplexing scheme is employed, in which a single radio front-end is used at both the Tx and Rx. A time-division multiplexing (TDM) is applied, where each antenna sequentially transmits with appropriate guard intervals inserted between transmissions. Although no additional SNR gain at the Rx can be achieved through simultaneous repetitive transmission from all Tx antennas, this approach simplifies maintenance and calibration due to its straightforward hardware design. In this study, the USRP X310 \cite{Ettus} is adopted, consisting of a motherboard and an RF daughterboard (UBX 160) capable of operating over a frequency range from 10~MHz to 6~GHz with a maximum passband bandwidth of 160~MHz. The USRP is widely used for rapid and cost-effective prototyping, leveraging the USRP Hardware Driver (UHD) provides APIs for advanced control via C/C++ and Python. 

%%%%%%%%%%%%%%%%%%%%%%%%%%%%%%%%%%%%%%%%%%%%%%%%%%%%%%%%%%%%%%%%%%%%
\begin{figure*}[t]
\centering
\subfigure[Block diagram.\label{fig:measurement_system}]{\includegraphics[width=.5\linewidth]{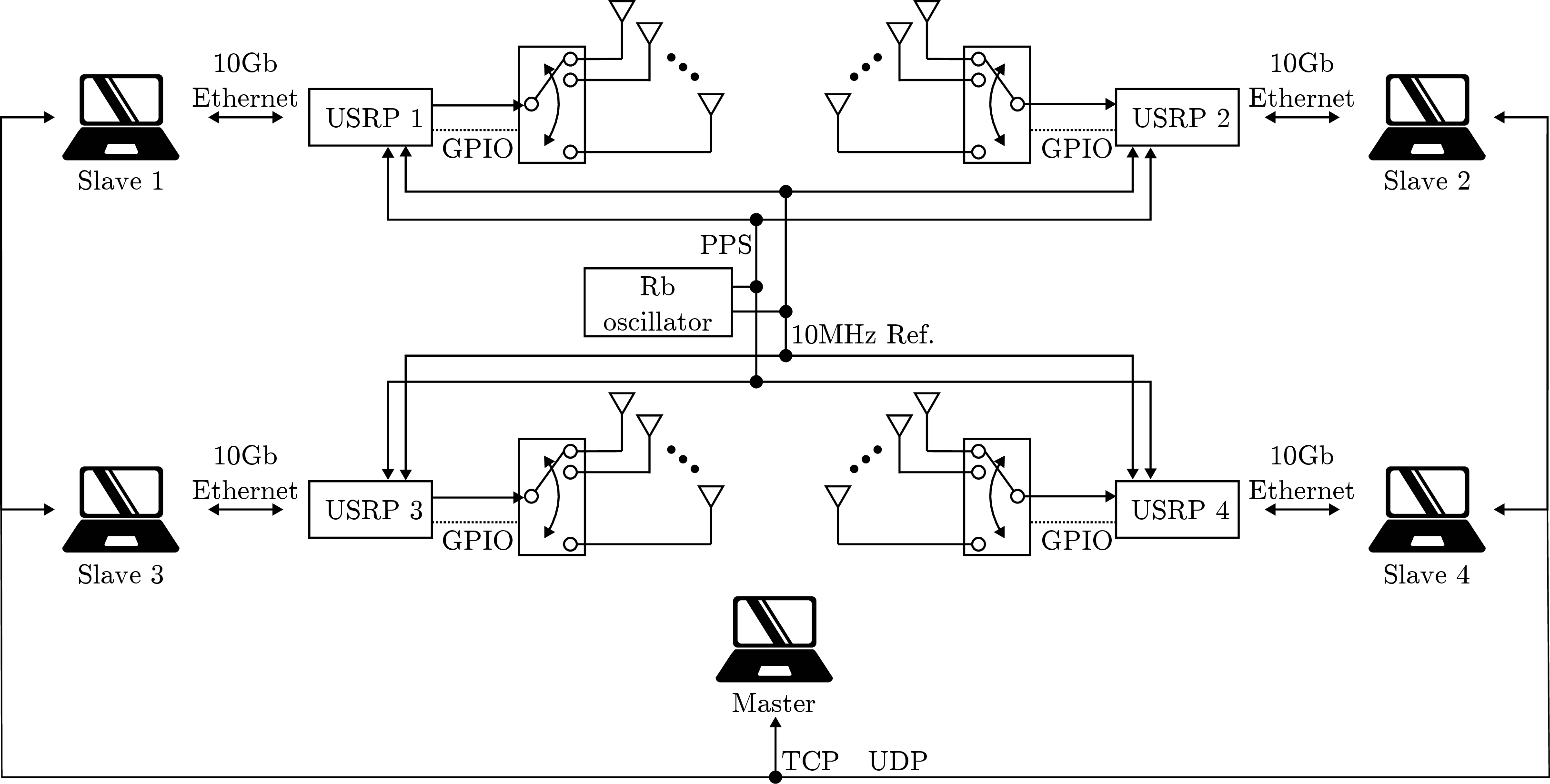}} \qquad
\subfigure[Photograph.\label{fig:measurement_equipment}]{\includegraphics[width=.33\linewidth]{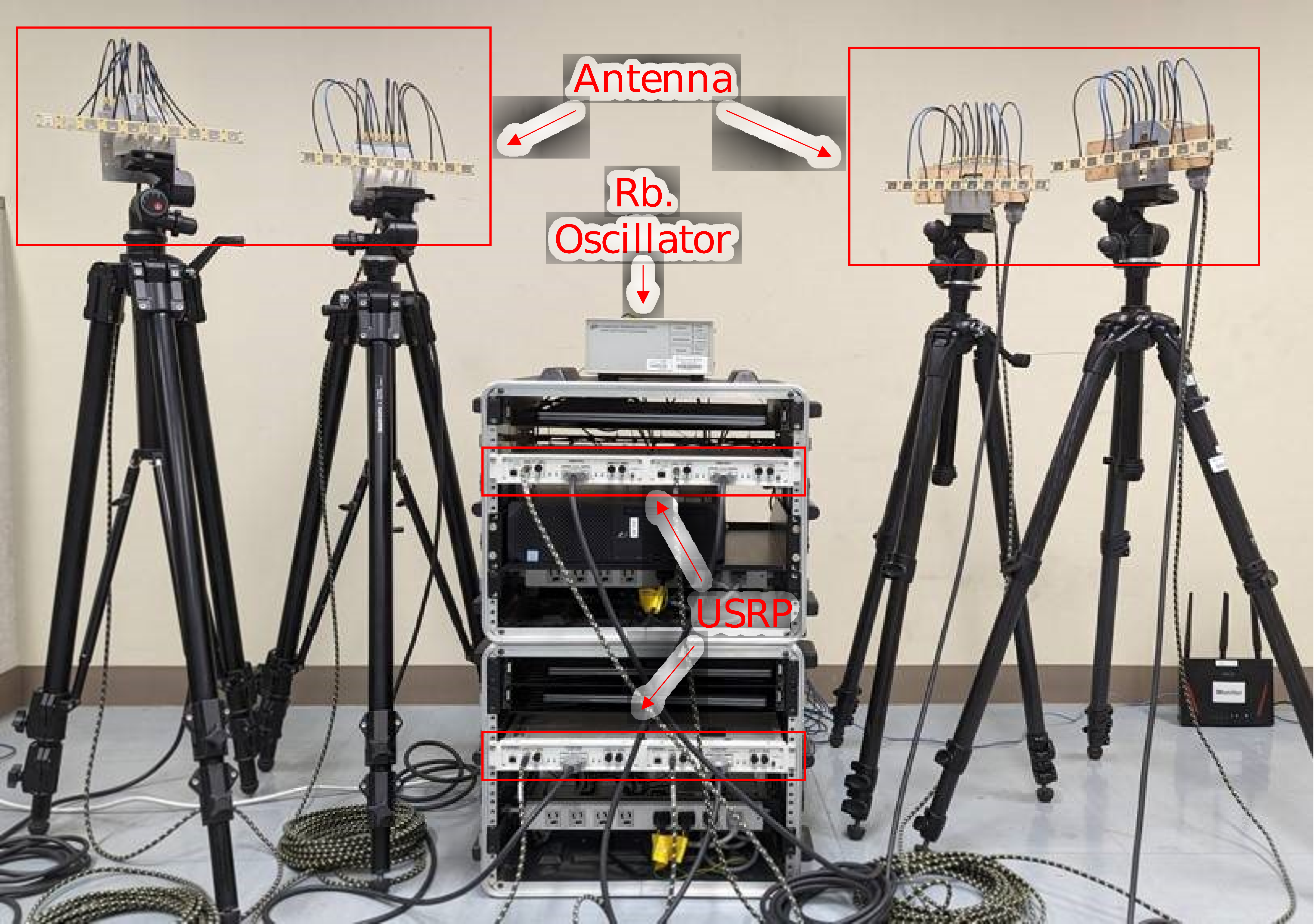}} 
\caption{Overview of prototype system for DAN-based DFL, consisting of four nodes equipped with USRP devices, controlled by a master-slave architecture for distributed channel measurements.}
\label{fig:measurement_system}
\end{figure*}
%%%%%%%%%%%%%%%%%%%%%%%%%%%%%%%%%%%%%%%%%%%%%%%%%%%%%%%%%%%%%%%%%%%%

An RF switch connected to the daughterboard's antenna port is used to switch the antennas to realize a switched antenna scheme. The RF switch (SP8T, SR-J030-8S, Universal Microwave Components Corp. \cite{UMCC}) covers a frequency range of $0.5$--$12.4$ GHz with a maximum insertion loss of $3.0$~dB. It offers high isolation of 60 dB ($0.5$--$6$ GHz) and a VSWR of 1.8:1 max in the on-state. Fast switching is achieved with a $40$ ns rise/fall time and $90$ ns on/off time. The USRP X310 features 15-pin general-purpose input/output (GPIO) ports which are directly controlled by the FPGA on the motherboard. The UHD API provides functions to operate the GPIO ports, and those are used to control the RF switch in the developed system. Fig.~\ref{fig:switching_sequence} shows an example control sequence diagram for $3\times 4$ MIMO antenna switching. The Tx transmits twice the number of sounding signals for the number of receiving antenna elements, then switches the Tx antenna, and the Rx switches antennas over a period twice the length of the sounding signal. This allows the capture of the sounding signal even with some time delay caused by propagation and RF switching. In this study, a $8\times 8$ MIMO scheme is implemented. The detailed measurement system specifications are shown in Table~\ref{tab:measurement_system}. Due to the roll-off characteristics of the DDC, which distorts both ends of the transfer function, the middle 128 points out of the 256 points of the obtained transfer function, excluding 64 points from both ends, are used. As a result, the passband signal bandwidth becomes 100 MHz although the sampling frequency is 200 MSa/s. A multitone signal, an unmodulated OFDM signal, is used as the sounding signal \cite{Kim_ScalableSounder}.

%%%%%%%%%%%%%%%%%%%%%%%%%%%%%%%%%%%%%%%%%%%%%%%%%%%%%%%%%%%%%%%%%%%%
\begin{figure*}[t]
\centering
\subfigure[Timing synchronization for transmission and reception among nodes. \label{fig:sync_image}]{\includegraphics[width=.55\linewidth]{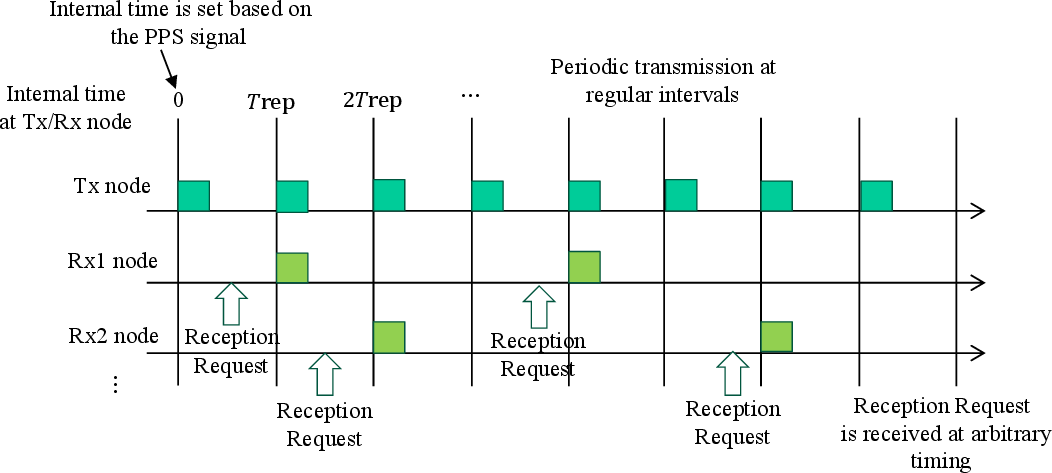}} \qquad
\subfigure[Six-link measurement procedure.\label{fig:multilink_image}]{\includegraphics[width=.35\linewidth]{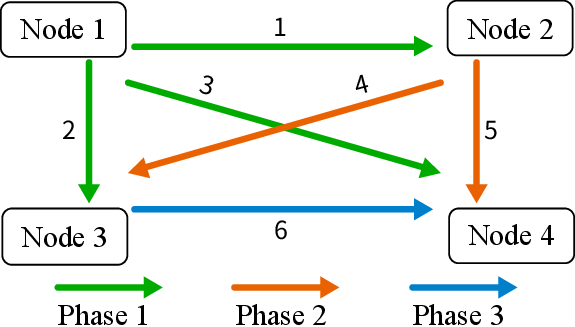}}
\caption{Multi-link channel measurement protocol of a DAN.\label{fig:DAN_protocol}}
\end{figure*}
%%%%%%%%%%%%%%%%%%%%%%%%%%%%%%%%%%%%%%%%%%%%%%%%%%%%%%%%%%%%%%%%%%%%

\subsubsection{Phase Stability}

Accurate measurement of phase information is crucial in MIMO channel sounding, as the coherence of received signals across antenna elements significantly impacts array signal processing \cite{Kim_ScalableSounder,Konishi_MLSounder}, even when a common local oscillator (LO) signal is shared among the antennas. A switched antenna MIMO channel sounder repeatedly transmits the same sounding signal while sequentially switching the Tx and Rx antennas to measure all channels. Therefore, phase drift during the measurement process can significantly degrade the accuracy of array-based signal processing. To assess this effect, the phase stability of the LO was evaluated.

The evaluation was conducted by directly connecting the Tx and Rx antenna ports through an attenuator, transmitting a sounding signal, and measuring the phase of the received signal while synchronizing the devices with a common 10 MHz reference signal (REF). The measurement results are presented in Fig.~\ref{fig:phase_stability}. The total effect of the Tx and Rx phase variation was calculated by averaging the phase across all subcarriers and then measuring the phase difference at each time sample relative to the first sample. The results showed that the maximum phase difference was approximately \Ang{3}, with a standard deviation of \Ang{0.62}. The result also indicates that the absolute phase difference during a single snapshot of the $8\times8$ MIMO measurement (approximately $164~\mu$s) was found to be approximately \Ang{0.4}--\Ang{0.6}. These results suggest that the impact of the switched antenna scheme is minimal and that the system's phase stability is sufficiently acceptable for array signal processing applications.

\subsection{Multi-Link Channel Acquisition Scheme}
\subsubsection{Configuration}
In this section, we extend the switched antenna MIMO channel sounder to a multi-link sounding system developed to efficiently measure radio channels in a DAN. Each node is equipped with an 8-element switched antenna array for half-duplex transmission and reception, enabling the measurement of CIRs for $8\times8$ MIMO channels between each pair of nodes. The system is designed to be scalable; in this study, it is configured as a six-link system comprising four nodes, since the number of links for four nodes is given by ${4 \choose 2}$. Fig.~\ref{fig:measurement_system} illustrates an overview of the measurement system. Each node contains a USRP device controlled by a dedicated slave PC. A master PC coordinates the overall operation by sending transmission and reception commands to the slave PCs over TCP connections. The received data from the slave PCs are then transmitted to the master PC using UDP.

\subsubsection{Timing Synchronization}
To achieve propagation channel measurements with physically separated multiple radio transceivers, carrier frequency, and transmission/reception timing must be synchronized among them. USRPs have 10 MHz REF and pulse-per-second (PPS) ports, which function as interfaces for synchronizing multiple devices. In this study, the REF and PPS signals generated by a rubidium oscillator are commonly input to each node's USRP, achieving frequency and timing synchronization between the nodes. The UHD API provides a function to set the USRP device time synchronized with the PPS. By using this function to reset the device time to 0 when communication between the USRP device and the host PC is established, the USRP device time is synchronized with the PPS signal. This synchronizes the internal time (elapsed time from PPS) between the nodes. Fig.~\ref{fig:sync_image} illustrates the concept of synchronized transmission and reception timing. According to this figure, the Tx transmits the sounding signal every $T_\mathrm{rep}$, where the interval is arbitrarily chosen to ensure stable operation ($T_\mathrm{rep}=100$~ms, herein). The Rx captures the signal from an arbitrary transmission cycle. The transmission and reception timing can be specified by setting the device time for transmission and reception initiation using functions provided by the UHD API.

\subsubsection{Multilink Measurement Protocol}
By the method described above, it is possible to measure the CIRs of the MIMO channels for each link (Tx-Rx node pair within a DAN). This is extended to realize multi-link measurement by appropriately switching the transmission or reception mode of each node. Fig.~\ref{fig:multilink_image} shows the Tx-Rx pairs for each link measurement. In this scheme, the measurement of all links is divided into three phases under the assumption of channel reciprocity. In the first phase, Node~1 is set as the Tx node, and measurements for Link~1 to Link~3 are conducted. In the second phase, Node~2 is set as the Tx node, and measurements for Link~4 to Link~5 are conducted. Finally, in the third phase, Node~3 is set as the Tx node, and the measurement for Link~6 is conducted.

In each phase, communication is carried out between the master PC and the slave PCs. To ensure that transmission has started correctly, the command to initiate reception at the Rx slave PC is issued only after the master PC receives a transmission start notification from the Tx slave PC. Upon receiving the transmission start command, the Tx node transmits continuously at the specified intervals. Conversely, upon receiving the reception request command, the Rx node captures the required samples in synchronization with the transmission timing. When multiple Rx nodes are present, the master PC broadcasts the reception request command to all Rx slave PCs to initiate reception simultaneously. After reception, the received data are transferred from each slave PC to the master PC via UDP. The master PC verifies the received data and subsequently initiates the transmission for the next phase. To further expedite the procedure, it is also possible to save the received data locally on each slave PC, bypassing the data transfer step, and proceed directly to the next phase. It should be noted that the target's moving speed must be sufficiently low to ensure that the complete multi-link MIMO channel measurement remains unaffected.

%%%%%%%%%%%%%%%%%%%%%%%%%%%%%%%%%%%%%%%%%%%%%%%%%%%%%%%%%%%%%%%%%%%%
\begin{figure*}[t]
\centering
\subfigure[Measurement setup with three types of targets: Absorber, Subject~A, and Subject~B.\label{fig:measurement_node_position}]{\includegraphics[width=.43\linewidth]{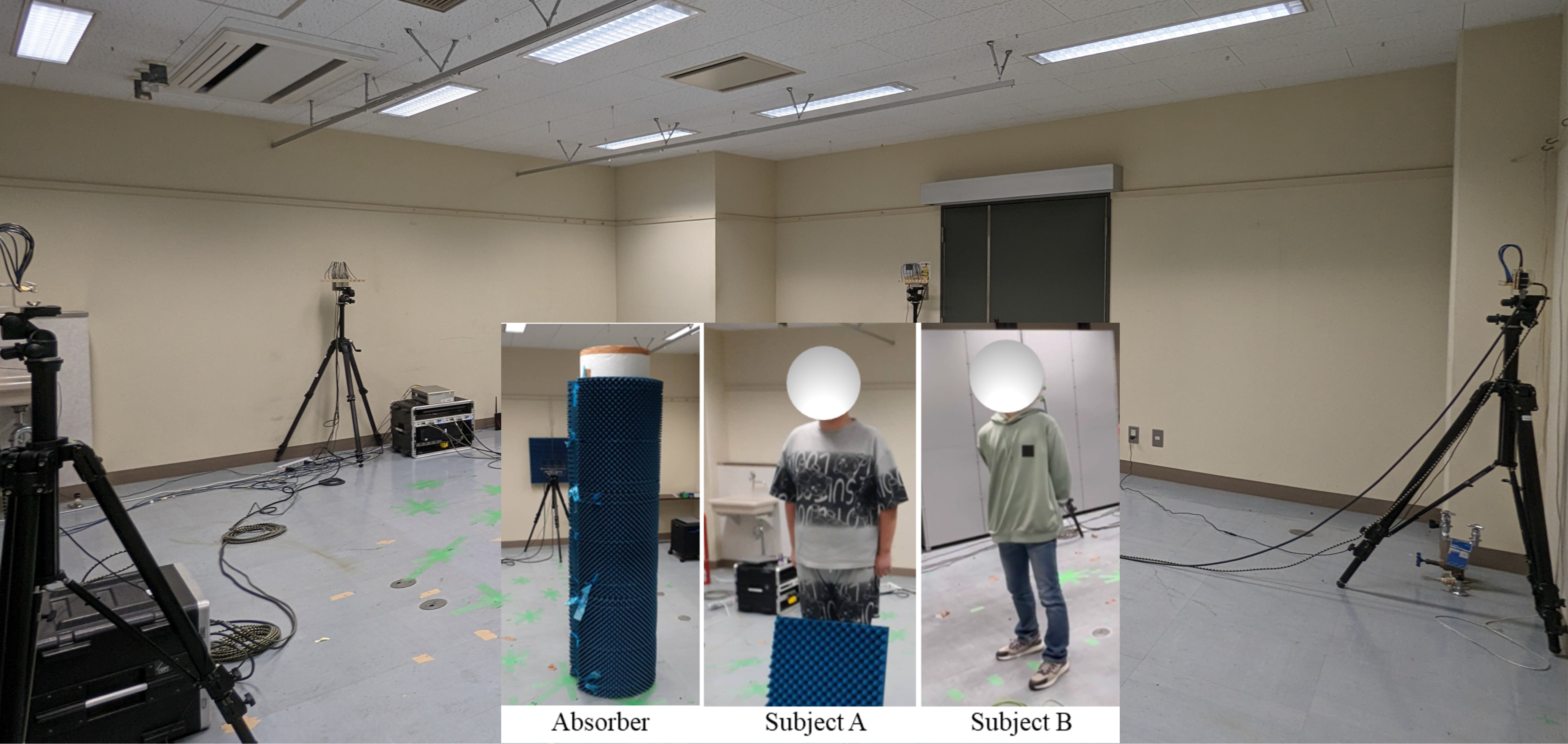}}  \quad
\subfigure[ULA with radiation patterns in the H-plane for individual patch antenna elements (4 dBi gain).\label{fig:ULA4_8G}]{\includegraphics[width=.52\linewidth]{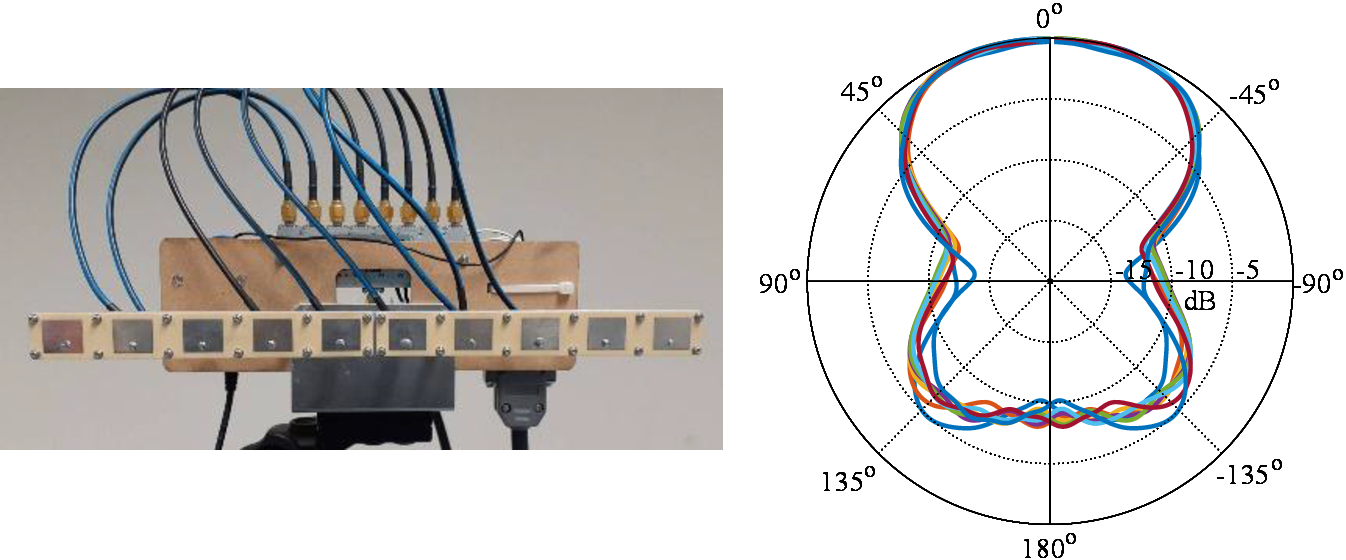}} 
\caption{Measurement setup. \label{fig:measurement_setup}}
\end{figure*}
%%%%%%%%%%%%%%%%%%%%%%%%%%%%%%%%%%%%%%%%%%%%%%%%%%%%%%%%%%%%%%%%%%%%

\section{Measurement-Based Evaluation} 
In this section, the localization accuracy is evaluated through measurements conducted using the multi-link MIMO channel sounding system developed for the four-node indoor DAN-based DFL system described in the previous section.

\subsection{Evaluation Setup}
The measurements were carried out in the actual room, which is identical to the evaluation environment depicted in Fig.\ref{fig:simulation_model}. In this setup, the anchor nodes indicated by red circular markers (Nodes 1--4) were used. Localization accuracy was evaluated using 25 distinct target positions, uniformly spaced at $0.5$~m intervals. The target positions are indicated by yellow square markers in Fig.~\ref{fig:simulation_model}. The measurement environment is illustrated in Fig.~\ref{fig:measurement_node_position}, which also depicts the three types of targets: a radio wave absorber and two human subjects. Consistent with the simulation setup, a cylindrical radio wave absorber with a diameter of $0.6$~m and a height of $2$~m was used to emulate a human target. Human Subject~A has a larger body size than Subject~B\footnote{Written informed consent was obtained from all participants prior to the measurements.}. Three types of targets were selected to evaluate their impact on actual shadowing effects, as the simulation assumed that signal paths obstructed by a target were fully shadowed. Each node was equipped with a ULA comprising 10 patch antenna elements, spaced at half-wavelength intervals, including 2 dummy elements, as shown in Fig.~\ref{fig:ULA4_8G}. The figure also presents the radiation patterns in the H-plane for each patch element, which has a gain of 4dBi and a half-power beamwidth (HPBW) of \Ang{90}. The Tx and Rx antennas for each link were positioned at the same height of $1.3$~m. 

\begin{figure}[t]
\centering
\includegraphics[width=\linewidth]{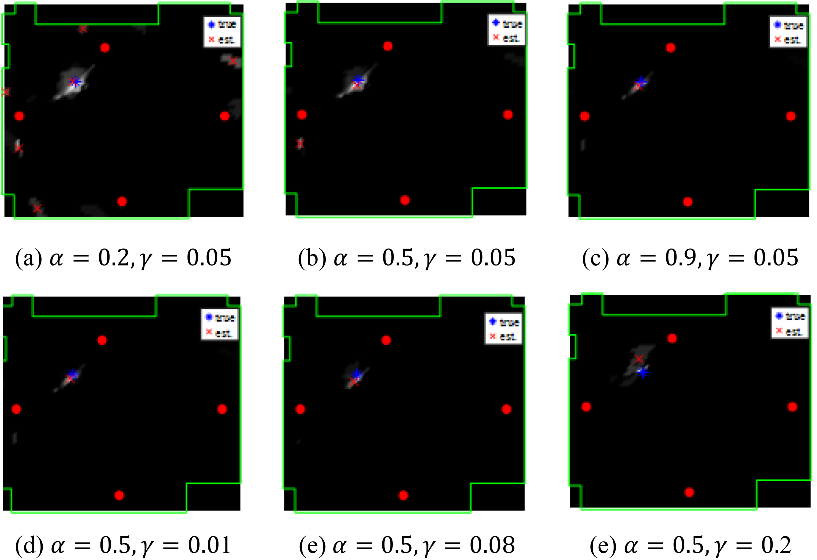}
\caption{Behavior depending on the sparsty $\alpha$ and path thickness $\gamma$.}
\label{fig:param_behavior}
\end{figure}

Channel measurements were conducted both in the absence of a target (baseline) and with a target present in each configuration. Post-processing was then performed to generate RTI images and estimate target positions. In the post-processing stage, the propagation paths computed by RT, $\Vect{\Omega}_{l,n_l}$, were utilized. The vector of RSS changes in~\eqref{eq:delta_y} is obtained through double-directional spatiotemporal beamforming using $\Vect{\Omega}_{l,n_l}$, applied to the measured MIMO channel vectors, as expressed in~\eqref{eq:delta_yln}. Thereafter, the same procedure used in the simulations in Sect.~III was applied to generate the images. The optimization parameters fitted to the measurement data are summarized in Table~\ref{tab:measurement_parameter}. As described in \cite{Access_Ikegami}, artifacts in the produced image decrease with an increase in the sparsity parameter $\alpha$. However, image quality is influenced not only by sparsity but also by the thickness $\gamma$ of the propagation path. Fig.~\ref{fig:param_behavior} illustrates the relationship between these two parameters, indicating that as $\alpha$ increases and $\gamma$ is appropriately tuned, artifacts are reduced and the estimated position aligns more closely with the true target. In this study, due to the limitations of heuristic parameter selection, Bayesian optimization \cite{Bayesian} was employed. This method balances exploration and exploitation to minimize localization error by reducing the discrepancy between the estimated and true positions, thereby determining the optimal parameters.

%%%%%%%%%%%%%%%%%%%%%%%%%%%%%%%%%%%%%%%%%%%%%%%%%%%%%%%%%%%%%%%%%%%%
\begin{table}[t]
\centering
\caption{RTI parameters}
\label{tab:measurement_parameter}
\begin{tabular}{c|c|c}
\hline
& Parameters                          & Values                         \\
\hline
\multirow{2}{*}{\begin{tabular}{c}RTI \end{tabular}} & voxel size                     & $0.1$ m                       \\
 & thickness, $\gamma$       & Bayesian optimization       \\ \hline 
\multirow{2}{*}{\begin{tabular}{c}Regularization \end{tabular}} & intensity, $\lambda$   & 5-fold cross-validation \\
 & sparsity, $\alpha$    & Bayesian optimization           \\ \hline
\multirow{2}{*}{\begin{tabular}{c}DBSCAN \end{tabular}} & $\varepsilon$                  & $0.5$                         \\
 & $N_\mathrm{minPts}$       & $3$                                  \\
\hline
\end{tabular}
\end{table}
%%%%%%%%%%%%%%%%%%%%%%%%%%%%%%%%%%%%%%%%%%%%%%%%%%%%%%%%%%%%%%%%%%%%

\subsection{Results}
\begin{figure*}[t]
\centering
\includegraphics[width=.955\linewidth]{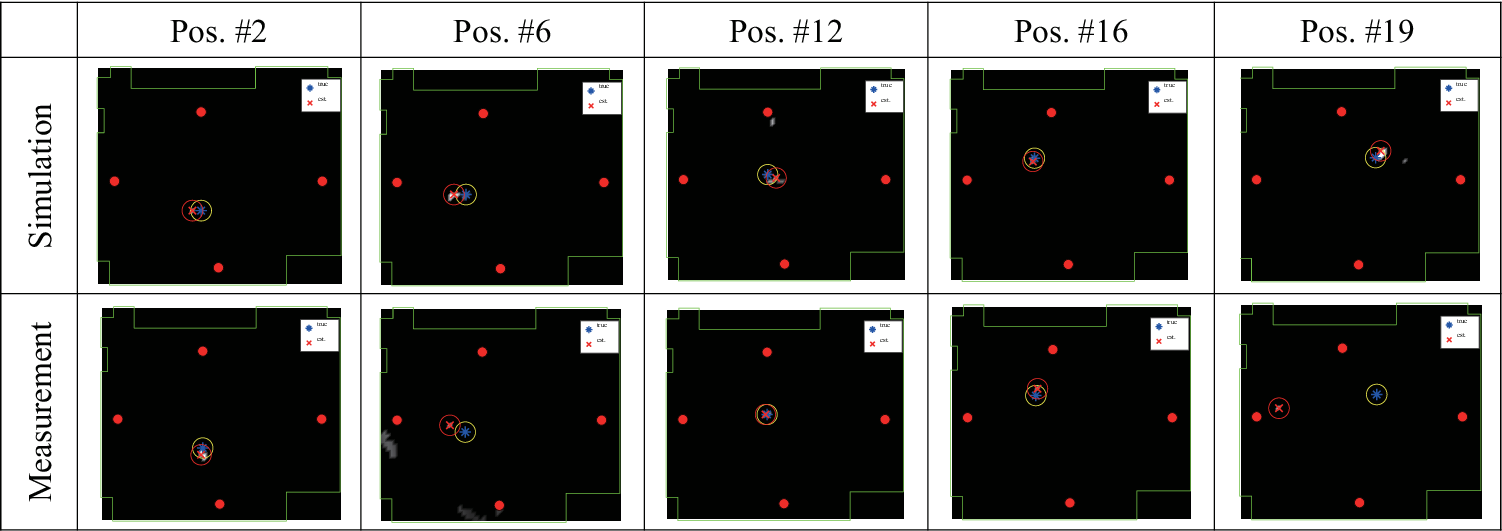}
\caption{Generated RTI images and localization results}
\label{fig:measurement_result}
\end{figure*}

Fig.~\ref{fig:measurement_result} presents five RTI-based localization results for positions 2, 6, 12, 16, and 19, obtained from both simulation and measurement, where the colored circles around the markers represent the target areas. Except for position 19, the estimated positions closely align with the true positions, often falling within the innermost target region, indicating high localization accuracy in both simulation and measurement. This consistency across different points reflects stable system performance. However, at position 19, a noticeable discrepancy between the estimated and true positions is observed. This degradation in accuracy is attributed to the reduced number of multipath components intersecting the target position, which increases localization ambiguity. Additionally, potential discrepancies between the RT results and the actual multipath propagation in the environment further exacerbate this uncertainty. These findings underscore the sensitivity of localization performance to multipath conditions and highlight the necessity of accurate propagation path estimation to ensure robust and reliable localization.

%%%%%%%%%%%%%%%%%%%%%%%%%%%%%%%%%%%%%%%%%%%%%%%%%%%%%%%%%%%%%%%%%%%%
%%%%%%%%%%%%%%%%%%%%%%%%%%%%%%%%%%%%%%%%%%%%%%%%%%%%%%%%%%%%%%%%%%%%
%%%%%%%%%%%%%%%%%%%%%%%%%%%%%%%%%%%%%%%%%%%%%%%%%%%%%%%%%%%%%%%%%%%%
%%%%%%%%%%%%%%%%%%%%%%%%%%%%%%%%%%%%%%%%%%%%%%%%%%%%%%%%%%%%%%%%%%%%
%%%%%%%%%%%%%%%%%%%%%%%%%%%%%%%%%%%%%%%%%%%%%%%%%%%%%%%%%%%%%%%%%%%%
\begin{figure*}[t]
\centering
\subfigure[Number of nodes (signle antenna per node); the subject is absorber.\label{fig:cdf_meas_Nodes_1elem_Absorber}]{\includegraphics[width=0.42\linewidth]{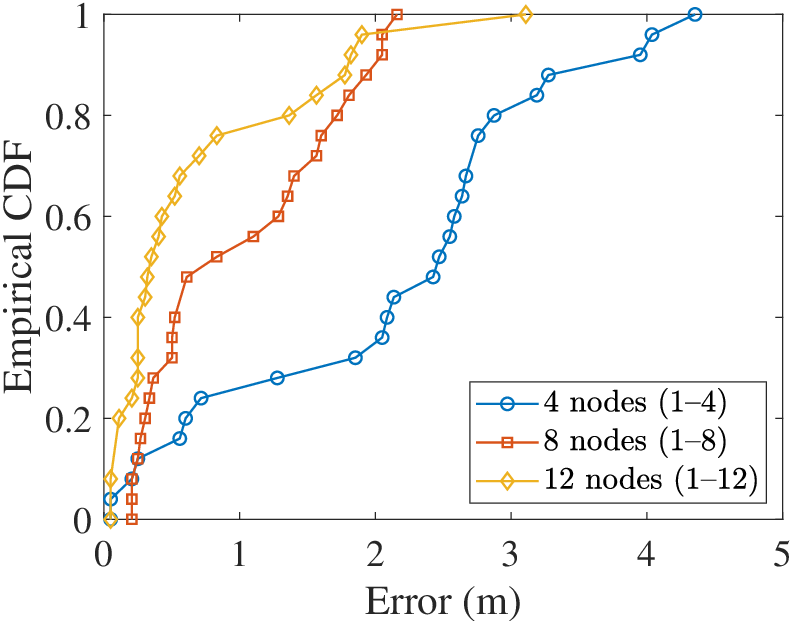}} \qquad
\subfigure[Different targets (four nodes with eight antennas per node). \label{fig:cdf_meas_subject_BAYS}]{\includegraphics[width=0.42\linewidth]{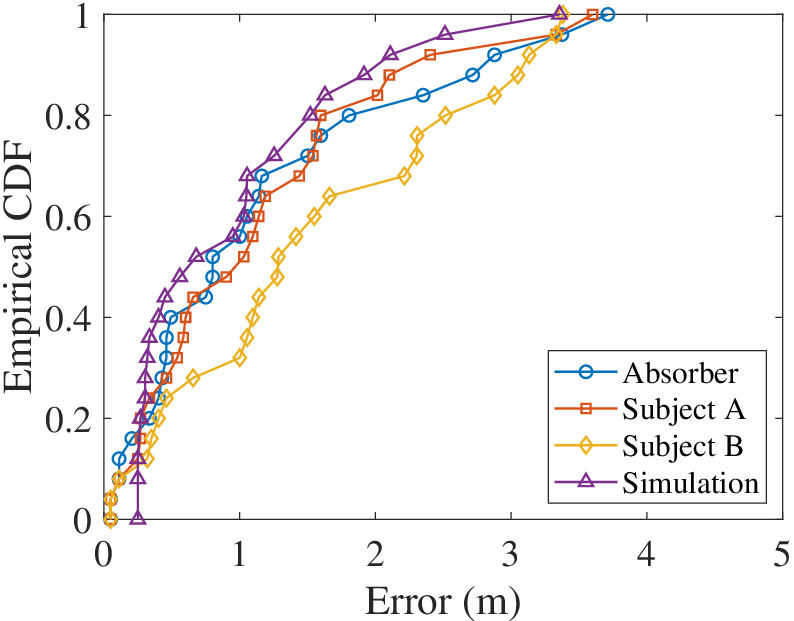}}
%\subfigure[Number of antennas (four nodes).\label{fig:cdf_Elems}]{\includegraphics[width=0.32\linewidth]{fig/cdf_Elems.eps}} \quad
\caption{Measurement results.}
\label{fig:MeasurementResults}
\end{figure*}
%%%%%%%%%%%%%%%%%%%%%%%%%%%%%%%%%%%%%%%%%%%%%%%%%%%%%%%%%%%%%%%%%%%%

\subsubsection{Number of nodes (single antenna)}
Fig.~\ref{fig:cdf_meas_Nodes_1elem_Absorber} shows the localization accuracy as a function of the number of nodes, where each node was equipped with a single antenna. The cylindrical radio wave absorber was used as a target. Due to installation constraints, the antennas were mounted on walls, which differs slightly from the simulation setup. Nonetheless, the measured results exhibit a similar trend to the simulation results shown in Fig.~\ref{fig:cdf_Nodes_1elem}. As previously discussed, an insufficient number of multipath components, particularly in configurations with 4 and 8 nodes, leads to a significant degradation in localization accuracy. In contrast, the configuration with 12 nodes provides acceptable localization performance.

\subsubsection{Target dependency}
The cylindrical radio wave absorber was utilized to ensure consistency with the simulation model and to support reproducibility. However, evaluating performance with actual human subjects remains essential, as multipath shadowing effects can vary significantly depending on body size and shape, thereby affecting system performance. Fig.~\ref{fig:cdf_meas_subject_BAYS} presents the localization accuracy of the four-node DAN configuration for three different targets: Absorber, Subject~A, and Subject~B. In this setup, each node is equipped with eight antenna elements, and the corresponding simulation result is included for comparison. As expected, the absorber scenario closely aligns with the simulation. Subject~A, characterized by a larger body size, introduces a moderate increase in localization error, whereas Subject~B, with a slimmer body, exhibits significant degradation in performance across the distribution. This discrepancy is likely due to differences in body size and the associated interaction with MPCs, which alters the shadowing loss of MPCs intersecting the target. 

\subsubsection{Discussion}
The presented results demonstrate that localization accuracy in DAN-based DFL systems is significantly affected by environmental conditions, node configuration, and target characteristics. A reduced number of nodes limits multipath diversity, leading to performance degradation, whereas an increased number enhances accuracy. Human-dependent multipath shadowing, varying with body size and shape, introduces substantial uncertainty that cannot be fully captured in simulations. Parameter optimization methods, including Bayesian optimization, proved effective in tuning key parameters such as sparsity and path thickness, thereby improving image quality and localization accuracy. These findings underscore the importance of accounting for both system design and human variability in practical applications.

\subsection{Multi-Target DFL}
The feasibility of multi-target DFL has been preliminarily evaluated through simulations in \cite{Access_Ikegami}. The results demonstrated that distinct MPCs enabled successful localization of multiple targets, although performance degraded when the targets were in close proximity due to interference and voxel ambiguity. In practice, the human body exhibits significant electromagnetic interactions, including partial transmission, diffraction, and multiple scattering effects, particularly when the target is located in close proximity to the transmitter or receiver. While simulation results may reasonably approximate actual performance when the separation between targets is sufficiently large, measurement-based validation remains essential for comprehensively assessing multi-target scenarios and accurately observing inter-target electromagnetic interactions. Alternatively, electromagnetic analysis using full-wave EM simulations and simplified EM models \cite{Rampa}, can provide valuable insights into body-induced propagation effects.

\section{Conclusion}
This paper proposed a novel multipath-RTI DFL framework based on DANs, aimed at enabling ISAC capabilities for future 6G RANs. The system leverages multi-link MIMO channel measurements and multipath-RTI to estimate the location of a passive target without requiring any radio device. A prototype platform operating in the sub-6 GHz band was developed using SDRs, incorporating a switched-antenna design to enable efficient MIMO channel sounding and a scalable multi-link architecture. The proposed framework introduces a novel method for weight matrix generation, which fully exploits MPCs as virtual anchor nodes, thereby enhancing spatial coverage. An Elastic Net-based sparse reconstruction technique was employed to solve the inverse imaging problem inherent to RTI. Additionally, Bayesian optimization was utilized to precisely tune critical parameters, such as sparsity and path thickness, leading to significant improvements in image quality and localization accuracy.

Extensive simulation and experimental evaluations demonstrated that the proposed method achieves sub-meter localization accuracy on average across various conditions. The results further revealed that performance deteriorates with limited node configurations and improves with increased the number of separable MPCs. Human-dependent shadowing capability, affected by body size and position, were identified as a key factor impacting localization accuracy, emphasizing the need for robust modeling in practical systems. 

While the DAN-based DFL framework introduces additional system-level requirements, such as inter-node synchronization and environmental awareness, these are aligned with the native capabilities of emerging 6G DAN and cell-free architectures, which are expected to incorporate synchronized operation and environment-awareness. The results demonstrated the effectiveness of the proposed approach and demonstrate its potential for scalable, accurate ISAC systems in next-generation wireless networks.

\end{document}